\documentclass{article}
\usepackage{spconf,amsmath,graphicx}

\usepackage{tabu}
\usepackage{multirow}
\usepackage[export]{adjustbox}
\title{SPECTRAL MODIFICATION BASED Data AUGMENTATION FOR IMPROVING END-to-END ASR for CHILDREN'S SPEECH}
\name{Vishwanath Pratap Singh $^{\dagger}$ \qquad Hardik Sailor
$^{\dagger}$ \qquad Supratik Bhattacharya $^{\star}$  \qquad Abhishek Pandey$^{\dagger}$}
\address{$^{\dagger}$ Samsung R\&D Institute, Bangalore\\
 $^{\star}$ Birla Institute of Technology \& Science, Pilani}

\begin{document}
%
\maketitle
\begin{abstract}
Training a robust Automatic Speech Recognition (ASR) system for children's speech recognition is a challenging task due to inherent differences in acoustic attributes of adult and child speech and scarcity of publicly available children's speech dataset.
In this paper, a novel segmental spectrum warping and perturbations in formant energy are introduced, to generate a children-like speech spectrum from that of an adult's speech spectrum. Then, this modified adult spectrum is used as augmented data to improve end-to-end ASR systems for children's speech recognition. 
The proposed data augmentation methods give 6.5\% and 6.1\% relative reduction in WER on children dev and test sets respectively,  compared to the vocal tract length perturbation (VTLP) baseline system trained on Librispeech 100 hours adult speech dataset. When children's speech data is added in training with Librispeech set, it gives a 3.7 \%  and 5.1\%  relative reduction in WER, compared to the VTLP baseline system.

\end{abstract}
\section{Introduction}
\label{sec:intro}
ASR has been improved significantly in recent years  due to advancements in machine learning techniques. 
However, the performance of such systems for children’s speech suffers from the large inter-speaker variability due to differing rates of growth, and intra-speaker variability due to undeveloped pronunciation skills, especially at very young ages (3-12 years) \cite{ref6}. Furthermore, the lack of adequate and appropriate children speech resources add to the challenge of designing a robust ASR system \cite{C3}. Improving children's speech recognition has substantial implications due to the growing number of applications in human-computer interaction, online education, computer-assisted language learning, speech-enabled smart toys, etc.

The performance of the ASR system, trained on the majority of adult speech data, degrades significantly when tested with children’s speech \cite{ref4,C2}. Differences in anatomy and language expressions are the major cause of  the degradation. Anatomically, underdeveloped vocal tract, shorter and lighter vocal cords of children lead to higher resonant and fundamental frequencies and greater spectral variability \cite{ref5,ref6}. 
Recently, there are various studies to improve children's speech recognition by incorporating children’s speech in training \cite{C5,C6,C7}, transfer learning \cite{TL1,TL2}, and multi-task learning with children vs adult speech classification as an auxiliary task \cite{MTL}.

In this paper, we propose novel spectral modification based data augmentation methods for improving children's speech recognition. In the first part of this paper, we study 
differences between adult vs. children speech spectrum due to underdeveloped vocal tract and uneven development of different cavities responsible for speech production. In the second part, we incorporate those changes in the adult's speech spectrum such that the modified adult speech spectrum is similar to the children's speech spectrum. Features computed using modified adult's spectrum are used as an augmented dataset for training. 

\section{Related Work}
\label{sec:relatedwork}
Due to the scarcity of children's speech data to train ASR models, data augmentation is one of the key research areas for children ASR. In ASR literature, there are various conventional data augmentation techniques that either modify speech signal or its time-frequency representation. Signal-level modifications are achieved using tempo and speed perturbations which change the speaking rate (fast/slow) and also pitch of the speech signal, respectively \cite{AudioPerturb}. Currently, most of the end-to-end ASR training pipelines include spectral augmentation using the SpecAugment technique that significantly reduces overfitting problems in neural network training \cite{SpecAugment}.

Voice-conversion (VC) and Text-to-Speech (TTS) synthesis are explored to deal with the limited availability of children's speech datasets. 
In \cite{GAN1, GAN3, GAN4}, adult speech signals are modified using a cycle consistent generative adversarial networks (GAN) to synthetically generate speech data with acoustic attributes similar to child speakers, and the synthetically generated speech is combined with a training set. Synthetic speech signals generated from children’s TTS model were added to ASR training to improve performance on children test cases in \cite{TTS2} and \cite{TTSchildASR}. The stochastic feature mapping (SFM) technique was also explored to transform out-of-domain adult data to children's speech data in \cite{fainberg16_interspeech}.

In \cite{VTLNchildren}, DNN was used to predict the frequency warping factors for vocal tract length normalization (VTLN) for improving hybrid DNN-HMM children's speech recognition system. In another work \cite{formantChild}, formant modification of children's speech spectrum is proposed using Linear Predicting Coding (LPC) approach \cite{LPspeech} to minimize the mismatch between train and test sets. Both of these methods use a uniform warping factor for all the formants. In related work, LPC-based spectral warping was applied selectively on vowel and non-vowel locations along with time scale modification in \cite{VowelLPC}. More recently, a novel fundamental frequency-based frequency warping technique was proposed which was shown to improve the performance of children ASR when added with other data augmentation techniques \cite{F0modify}.

However, in a previous study \cite{ref5}, it was found that due to the uneven development of different cavities responsible for speech production, the warping factor is not uniform across formants and it varies significantly for different age groups of children speakers. Fig. \ref{fig:fig3} (a) shows the mean scaling  factor (\%) for the first 3 formants, averaged across all the vowels of children speakers,  with respect to adult male formants. 
Also, existing spectrum warping methods \cite{VTLNchildren,formantChild,VowelLPC} do not consider spectral variabilities due to an underdeveloped vocal tract, such as varying formant’s bandwidth and formant’s energy with respect to the adult spectrum.
Generally, these methods are used to normalize the children’s speech or acoustic features to minimize the mismatch between train and test datasets but not as an augmentation.



\begin{figure}[htb]

\begin{minipage}[b]{1.0\linewidth}
  \centering
  \centerline{\includegraphics[width=9.0cm, height=3.3cm, keepaspectratio]{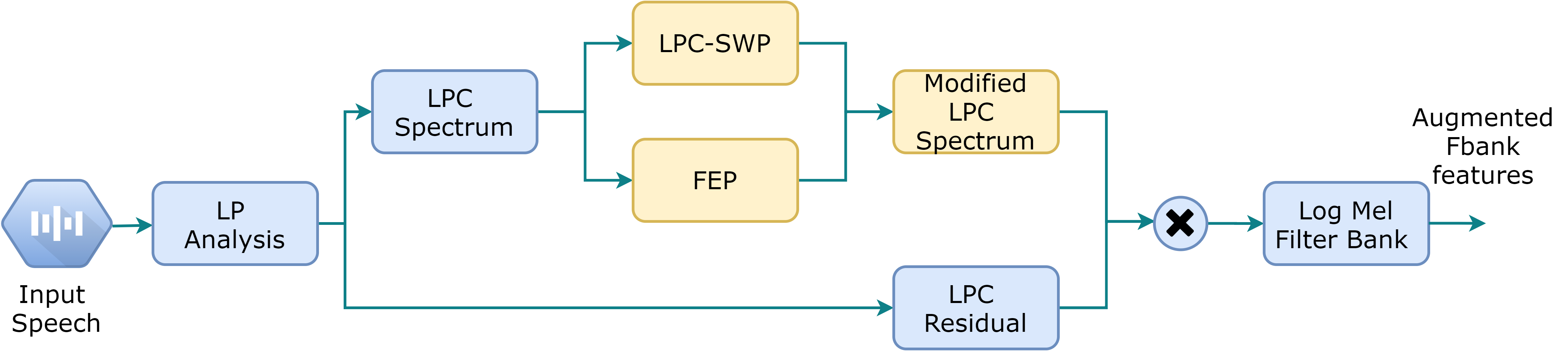}}
  \vspace{0.0 cm}
\end{minipage}
\caption{ A block diagram of the proposed LPC based data augmentation method.} 
\label{fig:fig0}
\end{figure}

\begin{figure*}[h]
\begin{minipage}[b]{0.31\linewidth}
  \centering
  \centerline{\includegraphics[width=5.75cm,height=4.4cm]{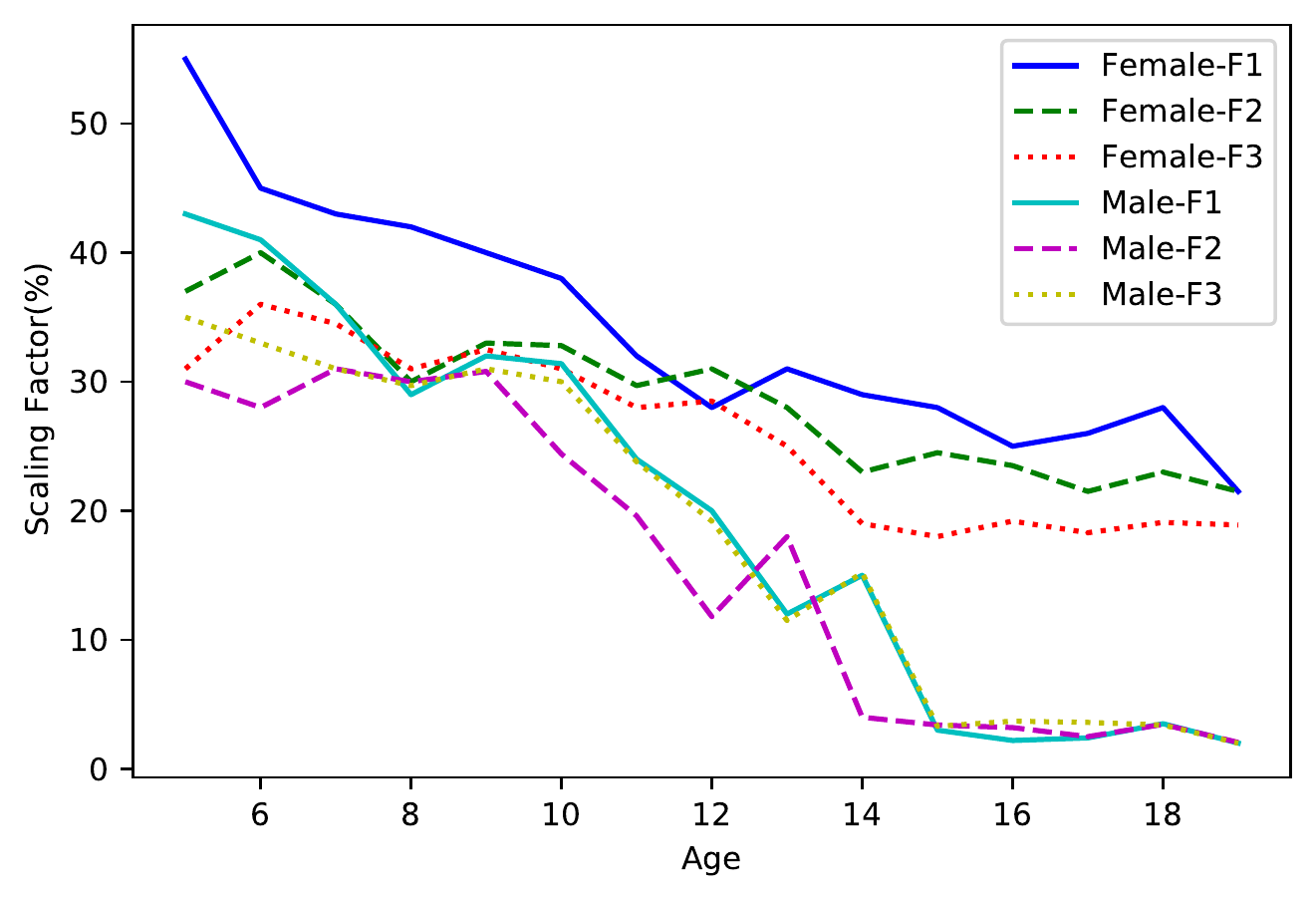}}
  \centerline{(a)} 
\end{minipage}
\hfill
\begin{minipage}[b]{0.31\linewidth}
  \centering
  \centerline{\includegraphics[width=5.75cm,height=4.4cm]{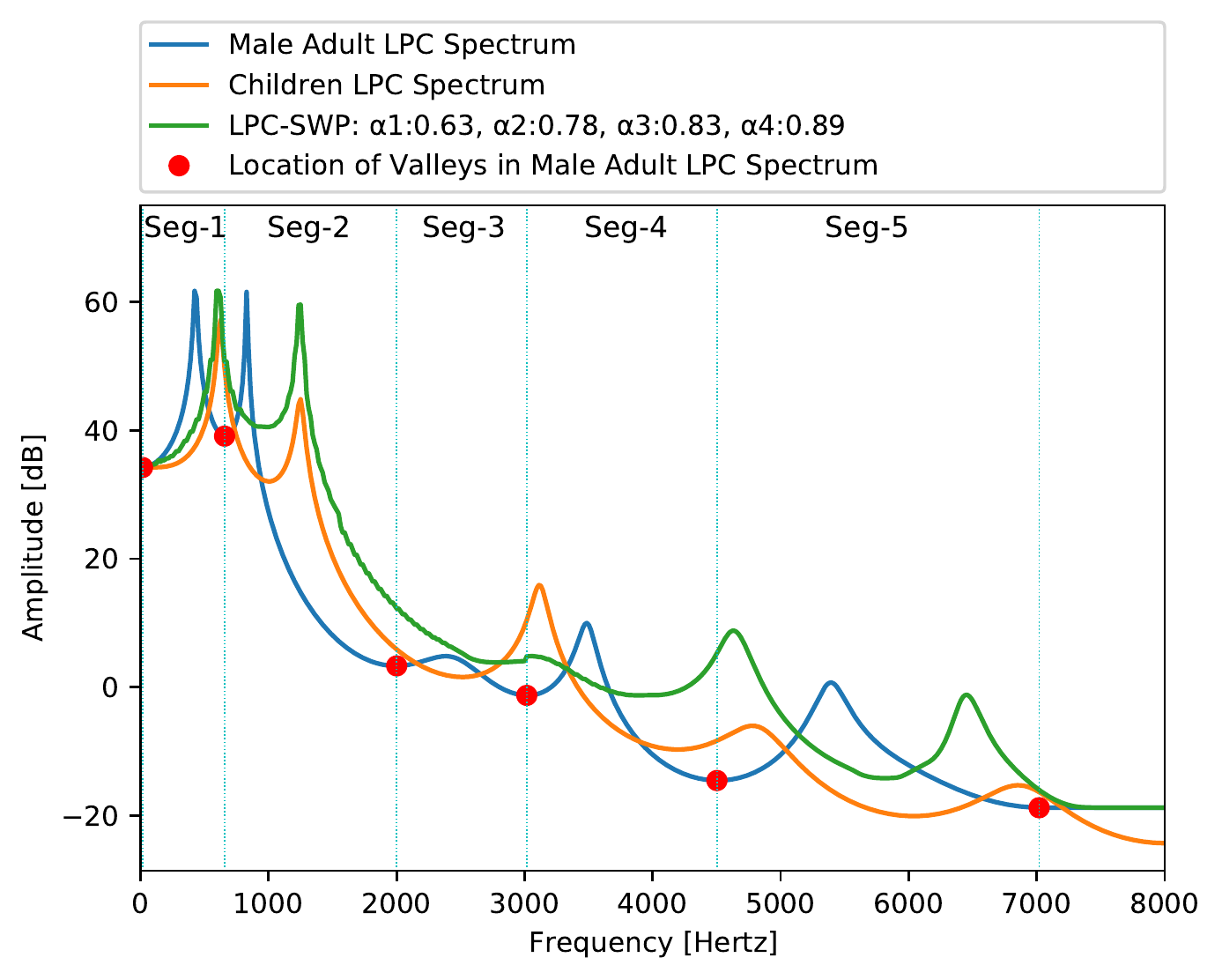}}
  \centerline{(b)} 
\end{minipage}
\hfill
\begin{minipage}[b]{0.31\linewidth}
  \centering
  \centerline{\includegraphics[width=5.75cm,height=4.4cm]{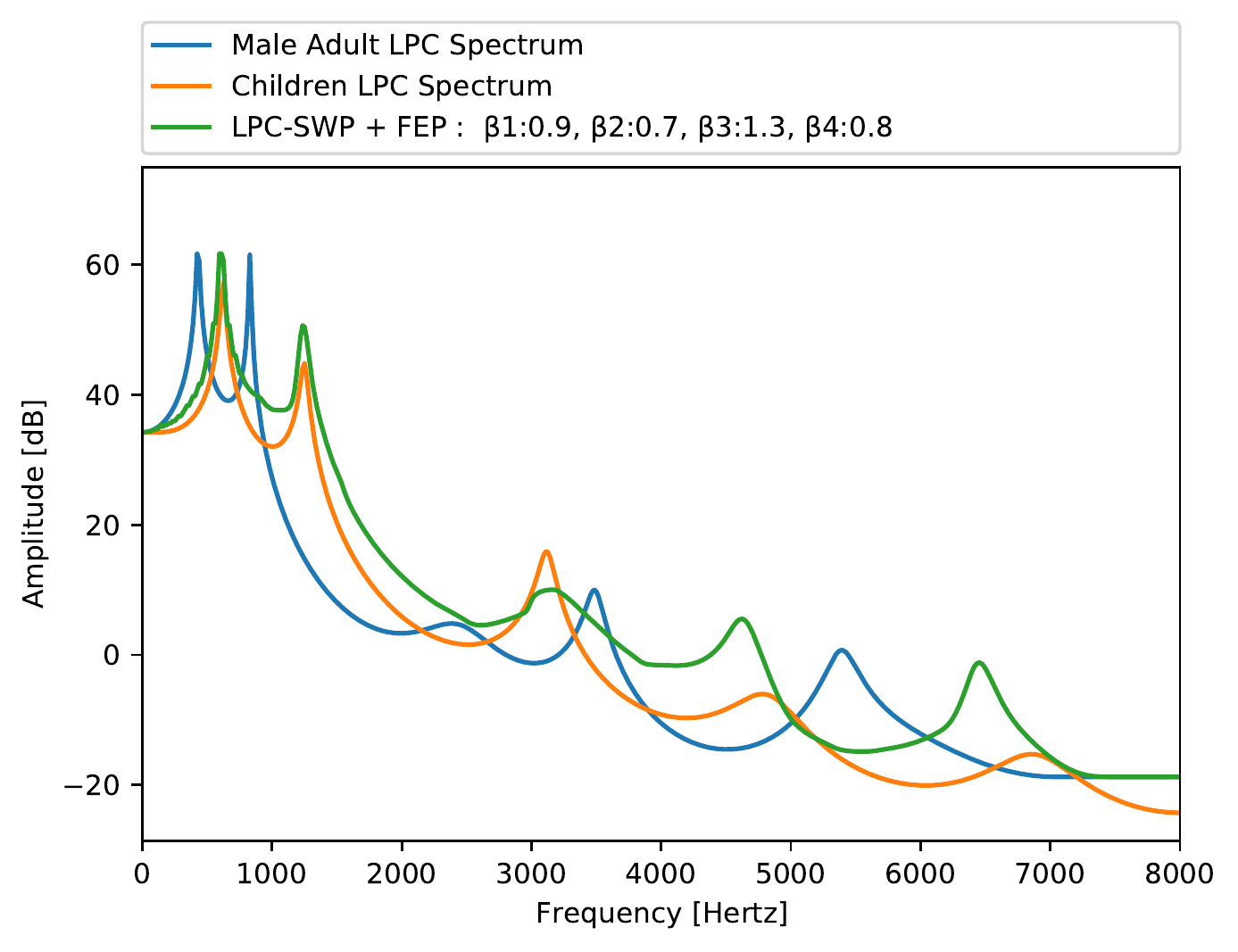}}
  \centerline{(c)} 
\end{minipage}
\caption{(a) Mean formant scaling factors (wrt adult) for first 3 formants of child speakers, averaged across vowels and subject to different age groups and genders. Data points are redrawn from \cite{ref5}, (b)-(c) LPC spectra of a child and adult speech, along with LPC-SWP and FEP based modified adult spectra,  for a vowel /O/ computed in 25 ms window. $\alpha_k$ and $\beta_k$ are frequency warping and energy scaling factors of the $k^{th}$ segment, respectively.}
\label{fig:fig3}
\end{figure*}


\section{Proposed Method}
\label{sec:proposed}
Differences between spectrum obtained using LPC technique of adult and child speakers are demonstrated in Fig. \ref{fig:fig3} (b). Three major differences can be observed between adult and child's LPC spectrum: 
\begin{enumerate} 
\item Scaling factors across formant frequencies of child's spectrum with respect to adults' spectrum are not uniform.
\item The first two and $4^{th}$ formant of child's spectrum have higher bandwidth than that of the adult's spectrum. This might be due to underdeveloped cavities of children's vocal tract responsible for the production of vowel /O/. It is also to be noted that $3^{rd}$ formant of the child's spectrum has wider bandwidth than that adult's spectrum. 
\item Energy distribution across formants is varying significantly. The first two formants of the adult speech spectrum have higher energy than that of child's speech spectrum, but the $3^{rd}$ formant of child's spectrum has significantly higher energy than that of adult speech spectrum. 
\end{enumerate}

In this paper, we incorporate these changes in the adult speech spectrum through LPC Segmental Warping Perturbation (LPC-SWP) 
and Formant's Energy Perturbation (FEP).  
We use this modified LPC spectrum to obtain the augmented children like filter-bank features for model training. Steps involved in the proposed method are shown through a schematic block diagram in Fig. \ref{fig:fig0}.  The range of warping factors and formant's energy scaling factors are presented in Table \ref{tab:warping}. Warping factors for vocal tract length perturbation (VTLP) and uniform LPC Wapring Perturbation (LPC-WP) based baseline systems are selected based on studies in \cite{Jaitly2013VocalTL}.  %
\begin{table}[th]
  \caption{Range of warping factors used in experiments} 
  \label{tab:warping}
  \centering
  \begin{tabular}{|c|c|} 
  \hline 
  Experiments & Warping Factors \\
  \hline\hline
  \multirow{1}{*}{VTLP} & $\alpha \in [0.9,1.1]$ \\ 
  \hline
  \multirow{1}{*}{LPC-WP} & $\alpha \in [0.9,1.1]$ \\
    \hline 
  \multirow{1}{*}{LPC-SWP (exp1)} & $\alpha_1 , \alpha_2, \alpha_3, \alpha_4 \in [0.9, 1.1]$ \\
    \hline  
 \multirow{1}{*}{LPC-SWP (exp2)} & $\alpha_1 , \alpha_2, \alpha_3, \alpha_4 \in [0.75, 1.0]$ \\
    \hline  
 \multirow{2}{*}{LPC-SWP (exp3)} & $\alpha_1 \in [0.6, 0.85]$, $\alpha_2 \in [0.7, 0.85]$, \\ & $\alpha_3\in [0.75, 0.95]$, $\alpha_4 \in [0.85, 1.0]$ \\
    \hline  
 \multirow{1}{*}{FEP} & $\beta_1 , \beta_2, \beta_3, \beta_4 \in [0.7, 1.3]$ \\
    \hline
 \end{tabular}
\end{table} 

\subsection{Segmental  Warping Perturbations (SWP)}
In this approach, we propose to augment the training
database by randomly generating a warp factor for each segment of the adult's speech spectrum. The range of warp factor for each segment is selected based on Fig. \ref{fig:fig3} (a) and a previous study \cite{ref5}. Here, the segments are defined by the region between the adjacent valleys. Different segments and effects of segmental warping perturbation of the LPC spectrum are shown in Fig. \ref{fig:fig3} (b). A random warp factor $\alpha_k$  is generated to warp the frequency axis of $k^{th}$ segment, such that
a frequency $f_k$ is mapped to a new frequency ${f_k}^{\prime}$ using
an approach similar to that proposed in \cite{Jaitly2013VocalTL,spkwarp}:

\[
    {f_k}^{\prime}= 
\begin{cases}
    \frac {f_k} {\alpha_k}, \hspace{1 cm} \text{if } {f_k }\leq max({f_k}^{max}, { F_{hi}}  \frac {max(1,\alpha_k)}  {\alpha_k} )\\
    \frac {S} {2} - \frac {\frac {S} {2} - \frac{ F_{hi}}{max(1,\alpha_k)}} {\frac {S} {2} - F_{hi}\frac {max(1,\alpha_k)}  {\alpha_k}  } (\frac{S}{2} - f_k), \hspace{1 cm}  \text{otherwise}
\end{cases}
\]
where, $S$ is the sampling frequency, $F_{hi}$ is the maximum frequency present in $4^{th}$ segment, ${f_k}^{max}$ is the maximum frequency present in $k^{th}$ segment. We limit the the number of segments up to 4 (i.e.,$k=1,2,3,4$) for our study.

It can be observed in Fig. \ref{fig:fig3} (b) that the first 4 formants of modified adult's speech spectrum are matched with that of the child's speech spectrum with a proper selection of warping factors for each segment. 
\subsection{Formant's Energy Perturbation (FEP)}
It is also to be noted in Fig. \ref{fig:fig3} (b) that after LPC-SWP, the formant's 
energy and it's bandwidth are still not matched. From Fig. \ref{fig:fig3} (c) it can be observed that with the proper scaling of each segment on the y-axis the formant energies of the modified adult speech spectrum are matched with the child's spectrum. To augment the training data, the magnitude of the $k^{th}$ segment is scaled with a random scaling factor $\beta_k$ chosen from a range of [0.7, 1.3] as mention in Table  \ref{tab:warping}. This range is decided based on analyzing the different vowels spectra across various age groups and their formant energies. 

\label{ssec:subhead}

\begin{table}[h]
  \caption{Dataset details}
  \label{tab:dataset}
  \centering
  \begin{tabular}{|c|c|c|c|c|} 
  \hline 
  Speaker & Dataset & Type  & \# Utt & Hours \\
  \hline\hline
  \multirow{5}{1.8cm}{Adult (Librispeech)} & TRAIN & train-clean-100 & 28539 & 100 \\ 
    \cline{2-5}
    & \multirow{2}{*}{VAL} & dev-clean & 2703  & 5.4 \\
    \cline{3-5}
    & & dev-other & 2864 & 5.3 \\
    \cline{2-5}
    & \multirow{2}{*}{TEST} & test-clean & 2620 & 5.4 \\
    \cline{3-5}
    & & test-other & 2939 & 5.1 \\
    \hline
    
    \multirow{3}{1.8cm}{Children (Internal)} & TRAIN & train-child-50 & 67062 & 50 \\  
    \cline{2-5} 
    & VAL & dev-child & 4000 & 3.0\\ 
    \cline{2-5} 
    & TEST & test-child & 4000 & 3.2 \\  
    \hline 
    
 \end{tabular}
\end{table}
\vspace{-0.4cm}
\section{Data Description}
\label{sec:data}
Experiments were performed using the publicly available 100 hrs subset of Librispeech corpus \cite{panayotov2015librispeech} and our in-house American English children speech data. Librispeech data contains the recordings of read speech from English audiobooks and speakers are well balance across adult males \& females. Children speech dataset also consists of read speech belonging to children's textbooks, storybooks, etc. and speakers age varies from 6 years to 15 years.  The details of both the datasets are given in Table \ref{tab:dataset}.

\begin{table*}[t]
  \caption{Results for ASR model with adult speech (\%WER)}
  \label{tab:results1}
  \centering
  \begin{tabular}{|c | c |  c | c |c |c |c|} 
 \hline
 Augmentation Method & dev-clean & dev-other & test-clean & test-other & dev-child & test-child \\ 
 \hline\hline
    SpecAugment & 10.3 & 26.9& 10.7&	27.1 &36.5&	36.1  \\ 
    SpecAugment + VTLP &9.8&25 &10.1 & 	25.9& 	35.4 & 	34.3  \\ 
    SpecAugment + LPC-WP &10& 24.4& 10.4& 25.2 &	36.7& 	35.6  \\
    SpecAugment + LPC-SWP (exp3) &9.6
&		25.6&	10.1&	26.1&	33.7& 	32.9 \\    
    SpecAugment + VTLP + LPC-SWP (exp1) & 9.7
 &  	24.7&  	10.2& 	25.8&  	34.9 & 	34.1 \\
    
    SpecAugment + VTLP + LPC-SWP (exp2) 
    &9.4& 24.6& 	9.9&25.5&34.6& 	33.5 \\

    SpecAugment + VTLP + LPC-SWP (exp3)  & 9.4
 & 	24.6& 	9.8& 25.4& 33.5& 	32.3	 \\

    SpecAugment + VTLP + LPC-SWP (exp3) + FEP &\textbf{9.3}&	\textbf{24.3}&\textbf{9.7}& \textbf{25.0} & \textbf{33.1} & \textbf{32.2} \\ 
    
 \hline
 \end{tabular}
\end{table*}
\begin{table*}[t]
 \caption{Results for ASR model with combined adult \& children speech (\% WER)}
  \label{tab:results2}
  \centering
  \begin{tabular}{|c | c |  c | c |c |c |c|} 
 \hline
 Augmentation Method & dev-clean & dev-other & test-clean & test-other & dev-child & test-child \\  
 \hline\hline
    SpecAugment & 9.8 & 25.0 &  10.3 & 25.8 & 8.3 & 7.7    \\
    SpecAugment + VTLP & 9.5
 & 	23.9 & 	9.6 &	25.1 & 	8.1& 	7.8 \\ 
   SpecAugment + VTLP + LPC-SWP(exp3) + FEP & \textbf{8.8}
 &	\textbf{22.9} &  	\textbf{9.2}	& \textbf{23.9} & 	\textbf{7.8}&	\textbf{7.4}  \\ 
 \hline
 \end{tabular}
\end{table*}

 \vspace{-0.4cm}
 
\section{Experiments}
\label{sec:pagestyle}
\subsection{Experimental Setup}
\label{ssec:expsetup}
All the end-to-end ASR experiments are performed using the ESPNet ASR toolkit \cite{watanabe2018espnet}. Our choice of model architecture is transformer \cite{Attention} and it consists of 12 encoder and 6 decoder layers. The multi-objective learning framework proposed in \cite{watanabe2017hybrid} is used to jointly optimize the CTC \& attention loss for 120 epochs. Mel filterbank features are computed using 25 ms window length with 10 ms of shift. For all the experiments, 5000 subwords units were generated using trained unigram on corresponding training text data.
In all of the experiments, SpecAugment \cite{SpecAugment} has been applied by default and other mentioned augmentation techniques are applied on top of SpecAugment. The training data size increased by 1-fold with the application of particular augmentation method (VTLP and LPC-based). For the decoding, we use the beam search algorithm with a beam size of 60 with a CTC weight of 0.4.
The decoding process did not include any language model in this paper.
\vspace{-0.2cm}
\subsection{Training ASR with adult speech}
\label{ssec:adultasr}
The first set of experiments aims to check the efficacy of proposed methods for the case if only the adult speech is available during the training time however the children's data is also available during the test time.
The experimental results of ASR models trained on 100 hours of Librispeech data are shown in Table \ref{tab:results1}. The models trained using SpecAugment and VTLP along with SpecAugment are treated as the baselines in this paper for comparison with our proposed methods. Applying VTLP along with SpecAugment decreased WER significantly for all the VAL \& TEST sets that make this a very strong baseline. 

Further, experiments are conducted with augmented features obtained from uniform LPC spectrum warping perturbation  (LPC-WP). LPC spectrum is obtained using $18^{th}$ order linear prediction filter with 25 ms frame duration and 10 ms frame shift. In our study, a model trained with VTLP features gives better performance compared to LPC-WP features for children's test cases. This might be due to the loss of information while obtaining the LPC spectrum from all-pole modeling of the vocal tract. However, since formants and spectrum segments can be located more efficiently using the LPC method, we decided to use the LPC spectrum for proposed spectrum augmentations. Interestingly, when SWP is applied to LPC spectrum (LPC-SWP (exp3) system), it gives a relative reduction of 4.8 and 4.0 \% in WER for children dev and test sets, respectively compared to VTLP baseline system. 

For the rest of the experiments, LPC-SWP is added along with VTLP features. Different warping factors were explored for LPC-SWP as shown in Table \ref{tab:warping}. The experiments were also conducted to justify why selecting the warping factor randomly for each spectrum segment is important than using uniform warping. Results in Table \ref{tab:results1} indicates that LPC-SWP (exp1) and LPC-SWP (exp2) did not perform well compared to LPC-SWP (exp3) where all 4 warping factors were selected randomly for each spectrum segment.
The best results with LPC-SWP (exp3) is obtained when the warping factor for each segment is chosen randomly from a range shown in Table \ref{tab:warping} which is motivated by the study presented in Fig. \ref{fig:fig3} (a). This indicates the significance of LPC-based segmental spectrum warping for children's ASR.
The last experiment  is conducted with features obtained from FEP, along with VTLP and LPC-SWP (exp3) features. This combination obtained reduced WER for all development and test cases compared to the baseline and other experiments.
\vspace{-0.3cm}
\subsection{Training ASR with adult and children speech}
\label{ssec:childasr}
 The experimental results for ASR models trained on the mixture of adult and children datasets are presented in Table \ref{tab:results2}. Adding children training data significantly reduced WER of both the children test sets compared to experiments in Table \ref{tab:results1}. 
 Adding VTLP further reduced WER for the adult test sets, however, there is no consistent reduction in WER for children's test sets. 
 Further adding SWP (exp3) and FEP based augmentation gives significant improvement over the VTLP baseline for children as well as adult speakers.
 Specifically, this feature combination gives a relative reduction of 3.7\% and 5.1\% in WER for dev-child and test-child test sets, respectively compared to the VTLP baseline.
 \vspace{-0.1cm}
\section{Conclusion}
In this paper, we introduce two new data augmentation techniques named LPC-SWP and FWP to improve children's speech recognition. These methods are based on the LPC spectrum and motivated by children's speech acoustics variability compared to adult speech especially the formant location and their energy distribution. Two sets of experiments were done one with only adult speech and another with a mix of adult \& children speech. For both experiments, our proposed data augmentation combination approach beats the baseline results significantly. In the future, it will be interesting to investigate the effect of formant's bandwidth perturbation and study the scope of LP residual augmentation for children's speech recognition task.

\newpage

\bibliographystyle{IEEEbib}
\small
\bibliography{strings,refs}

\end{document}